# Intensities of the Raman bands in the low-frequency spectra of DNA with light and heavy counterions

**S.M. Perepelytsya, S.N. Volkov**

*Bogolyubov Institute for Theoretical Physics, NAS of Ukraine, 14-b Metrologichna St., Kiev 03680, Ukraine;
e-mail: perepelytsya@bitp.kiev.ua; snvolkov@bitp.kiev.ua*

The approach for calculation of the mode intensities of DNA conformational vibrations in the Raman spectra is developed. It is based on the valence-optic theory and the model for description of conformational vibrations of DNA with counterions. The calculations for Na- and Cs-DNA low-frequency Raman spectra show that the vibrations of DNA backbone chains near 15 cm$^{-1}$ have the greatest intensity. In the spectrum of Na-DNA at frequency range upper than 40 cm$^{-1}$ the modes of H-bond stretching in base pairs have the greatest intensities, while the modes of ion-phosphate vibrations have the lowest intensity. In Cs-DNA spectra at this frequency range the mode of ion-phosphate vibrations is prominent. Its intensity is much higher than the intensities of Na-DNA modes of this spectra range. Other modes of Cs-DNA have much lower intensities than in the case of Na-DNA. The comparison of our calculations with the experimental data shows that developed approach gives the understanding of the sensitivity of DNA low-frequency Raman bands to the neutralization of the double helix by light and heavy counterions.

## 1. Introduction

The DNA double helix is stabilized by the metal counterions that neutralize the negatively charged phosphate groups, reducing electrostatic repulsion between opposite strands of macromolecule backbone [1]. Tethering to DNA phosphate groups, the counterions determine the structural and dynamical properties of the double helix [1-5]. Therefore, the study of localization of counterions and their dynamics in the double helix are important for understanding of the mechanisms of DNA biological functioning.

In water solutions the counterions and water molecules form an ion-hydrate shell around the DNA double helix. The thickness of this shell is about 7Å [6-8]. Inside the ion-hydrated shell the counterions spend part of the time in complex with DNA atomic groups and part of the time in free state. The molecular dynamics simulations [9-11] show that they are localized near the phosphate groups during about 0,1-1 ns that is much higher than the period of DNA conformational vibrations (~1 ps) [12-14]. In our previous works the approach for description of conformational dynamics of DNA double helix with counterions has been developed [15-17]. In framework of this approach the counterions have been considered tethered to the phosphate groups of macromolecule backbone forming a regular dynamical structure along the DNA double helix. Due to the regularity of the DNA backbone such a structure of counterions and phosphate groups may be considered as an ionic lattice that is characterized by the specific ion-phosphate vibrations. The vibrations of DNA ion-phosphate lattice should be observed in experimental spectra and may be used for the determination of counterion type in DNA structure.

In our previous works [15-17] to determine the frequency range of DNA ion-phosphate vibrations the dynamics of DNA with alkali metal counterions has been studied. The obtained modes of ion-phosphate vibrations were in the low-frequency spectra range. As it is known [12-14] in this spectra range there are modes of DNA conformational vibrations that characterize the vibrations of the double helix backbone (near 20 cm$^{-1}$), stretching of H-bonds in the base pairs, and intranucleoside vibrations (60 – 110 cm$^{-1}$). The calculations of the frequencies of conformational vibrations for DNA with Na$^+$, K$^+$, Rb$^+$, and Cs$^+$ counterions [15-17] show that the values of ion-phosphate frequencies are at the spectra range from 90 to 200 cm$^{-1}$ depending on counterion type. The frequency of ion-phosphate vibrations decreases as counterions mass increases. The character of internal dynamics of the double helix changes as counterion mass increases. For example, the modes of ion-phosphate vibrations of Cs-DNA are characterized by high amplitudes of intranucleoside vibrations and H-bond stretching in the base pairs, while the amplitudes of Na-DNA for the same displacements are small. So, the obtained results show that the dynamics of DNA conformational vibrations is sensitive to the type of counterions that form the ion-phosphate lattice.

The dependence of DNA conformational vibrations on counterion type is reflected in the experimental spectra. For example, in the infrared absorption spectra of the dry films of polynucleotides the mode depended on counterion type has been observed [18]. The frequency of this mode decreases as counterion mass increases the same as the calculated frequency of the mode of ion-phosphate vibrations [15-17]. To find the mode of ion-phosphate vibrations in our previous work [19] the Raman spectra of DNA water solutions with $Na^+$ and $Cs^+$ counterion have been studied. The obtained spectra show that under the substitution of $Na^+$ counterions for $Cs^+$ the intensity of the band near 100 cm$^{-1}$ increases. According to our calculations [15-17] such intensity increase arises due to the modes of DNA ion-phosphate vibrations. But to confirm the existence of the ion-phosphate mode in the DNA low-frequency Raman spectra the intensities of the modes of conformational vibrations of the double helix with counterions should be calculated.

In the present work the approach for determination of the intensities of the modes of DNA conformational vibrations in the Raman spectra is developed. This approach is based on the model of conformational vibrations of DNA double helix with counterions [17] and the valence-optic theory for derivatives of the polarizability with respect to the normal coordinates [20]. Using the developed approach the frequencies, amplitudes, and the Raman intensities are calculated for the conformational vibrations of DNA with $Na^+$ and $Cs^+$ counterions and the interpretation of the experimental spectra is given.

## 2. Approach for investigation of the intensities of DNA conformational vibrations in the Raman spectra

The intensities of the modes of molecular vibrations may be studied by the methods of quantum mechanics and semi-classical methods. But recent calculation results [21-23] show that semi-classical theory is the most fruitful for studying the macromolecular systems [20]. That is because the use of quantum mechanical theory for the calculations of such objects as DNA at the present is impossible due to high number of atoms in the monomer link of the double helix and due to the necessity of consideration of surrounded water molecules and counterions. Therefore, the semi-classical theory is assumed as the basis of our approach for the determination of intensities of the modes of DNA conformational vibrations in the Raman spectra.

In framework of semi-classical theory all molecular vibrations are classified as valence and deformation vibrations. The valence vibrations are characterized by changes of length of chemical bonds, and the deformation vibrations are characterized by changes of angles. Since the conformational vibrations of DNA are characterized by the changes of valence and torsion angles [12-14], they may be referred to deformation vibrations. In this connection all rules that are right for the deformation vibrations should be also right for the conformational vibrations. Particularly, the intensity of the modes of conformational vibrations should be determined by the polarizabilities of the chemical bonds that change the direction during their vibrations.

It is clear that the analysis of intensities will be essentially simpler if we take into consideration the character of vibrations of structural elements in the monomer link for each mode of DNA conformational vibrations. The four-mass model with counterions [15-17] takes into consideration the peculiarities of motion of the double helix atomic groups with counterions and describes well the low-frequency spectra of DNA. Therefore, it is used in the present work for the analysis of intensities of the double helix conformational vibrations.

The model is shown in the figure 1. The coordinates of the model are determined in the reference frame, where the axis $Y$ is directed parallel to the line connecting the atoms $C_{3'}$ of nucleotides of different chains of the double helix, the axis $Z$ is directed along the double helix, and the axis $X$ is directed to DNA major groove. Such reference frame is connected with the structure of the double helix, and in the present work it is considered as molecular reference frame.

In framework of the model of DNA conformational vibrations the motions of phosphates, nucleic bases, and desoxyribose rings as whole are considered. The nucleic bases and desoxyribose rings vibrate jointly as the physical pendulums with respect to the strands of macromolecule backbone and also with respect to each other in nucleoside (intranucleoside mobility). The physical pendulums are characterized by the reduced length $l$. The rotational vibrations of pendulum-nucleosides with respect to the phosphate groups are described by deviations $\theta_1$ and $\theta_2$ from the equilibrium angle $\theta_0$ that describe their position in the plane $XOY$ of complementary base pair of DNA (Fig. 1). The intranucleoside mobility that is conditioned by the conformational mobility of desoxyribose is described by changes of the reduced lengths of pendulum-nucleosides $\rho_1$ and $\rho_2$ (Fig. 1). The displacements of nucleotides (nucleoside + phosphate) are described by coordinates $X_1$, $Y_1$, $X_2$, and $Y_2$. The stretching of H-bonds is determined the as the distance between mass centers [17]: $\delta \approx Y_1 + Y_2 + l(\theta_1 + \theta_2) + \cos\theta_0 (\rho_1 + \rho_2)$.

The same as in our previous work [17] we consider that the phosphate groups are neutralized by counterions of the solution and form the dynamical ion-phosphate lattice (Fig. 1). The vibrations of counterions with respect to the phosphate groups are described by coordinates $\xi_1$ and $\xi_2$ for the first and the second strands of the double helix, respectively. In contrast to the earlier works [15-17], here the vibrations of counterions are considered independently on the vibrations of nucleosides. That is because the correlation between counterion and nucleoside vibrations is rather complex and its consideration is not needed for the solution of problem formulated in this work.

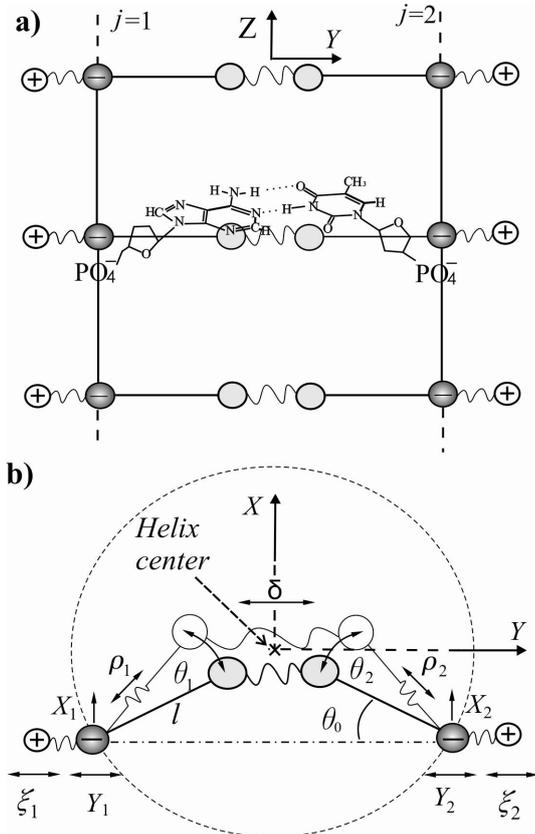

**Fig. 1.** Four-mass model with counterions. (a) Double chain of nucleosides, phosphates, and counterions. (b) Displacements of structural elements of the monomer link in the plane orthogonal to the helical axis. The coordinates $\xi$, $\theta$, $\rho$, $X$, and $Y$ describe the ion-phosphate vibrations, pendulum-nucleoside vibrations, intranucleoside vibrations, vibrations of phosphate group along X and Y axis, respectively. $l$ is the reduced length of nucleoside. $\theta_0$ is the equilibrium angle.

The motions of structural elements of the model are considered in the plane orthogonal to the helical axis because the longitudinal vibrations of the structural elements of macromolecule occur at higher frequency range [12-14]. The dependence of the parameters of the model on heterogeneity of nucleosides is not considered in the present work, since for the determination of relative intensities for the modes of DNA conformational vibrations it is no essential.

The model gives 7 modes of DNA conformational vibrations (see the works [16,17]). Two modes $\omega_B^+$ and $\omega_B^-$ describe the symmetric and antisymmetric vibrations of nucleosides as a pendulums. Two modes $\omega_H^+$ and $\omega_{HS}^+$ characterize symmetric vibrations that are accompanied by the stretching of H-bonds and deformation of the desoxyribose rings. One mode of antisymmetric vibrations $\omega_S^-$ describes the intranucleoside vibrations. Two modes of symmetric and antisymmetric ion-phosphate vibrations $\omega_{Ion}^+$ and $\omega_{Ion}^-$ characterize the vibrations of counterions with respect to the phosphate group.

The analysis of the model show that the amplitudes of vibrations for the nucleotides of the first and the second strands of the double helix are equal in the case of symmetric vibrational modes. In the case of antisymmetric modes the amplitudes have different phase of vibrations (the sign "+" or "-"). Therefore, the amplitudes of vibrations in the first and in the second nucleosides are interrelated as follows:

$$\begin{cases} \tilde{q}_1 = \tilde{q}_2 \text{ for } \omega_{Ion}^+, \omega_H^+, \omega_{HS}^+, \omega_B^+; \\ \tilde{q}_1 = -\tilde{q}_2 \text{ for } \omega_{Ion}^-, \omega_S^-, \omega_B^-; \end{cases} \quad (1)$$

where $\tilde{q}_1 = \{\tilde{X}_1, \tilde{Y}_1, \tilde{\theta}_1, \tilde{\rho}_1, \tilde{\xi}_1\}$, $\tilde{q}_2 = \{\tilde{X}_2, \tilde{Y}_2, \tilde{\theta}_2, \tilde{\rho}_2, \tilde{\xi}_2\}$ are the amplitudes of vibrations for the first and the second nucleosides, respectively.

## 3. Calculation method

According to semi classical theory [20] the intensity of some mode of molecular vibrations that is observed in the Stokes part of Raman spectra at the right angle geometry is determined as follows:

$$J_m = J_0 \frac{2^8 \pi^5 (\nu_0 - \nu_m)^4}{9 c^4} (5A_m^2 + 13B_m^2) Q_{0m}^2 \left(1 - e^{-\frac{h\nu_m}{kT}}\right)^{-1}, \quad (2)$$

where $J_0$ and $\nu_0$ are the intensity and the frequency of incident light, $\nu_m$ and $Q_{0m}$ are the frequency and amplitude of molecule normal vibration, $T$ is the temperature, $h$ is the Plank constant, $k$ is the Boltzmann constant. The index $m$ designs the mode of normal vibration. The constants $A_m$ and $B_m$ are the trace and anisotropy of tensor of derivatives of polarizability with respect to the coordinates of normal vibrations:

$$A_m = \sum_\lambda \frac{\partial \alpha_{\lambda\lambda}}{\partial Q_m}, \quad B_m = \sqrt{\frac{3}{2} \sum_{\lambda,\mu} \left(\frac{\partial \alpha_{\lambda\mu}}{\partial Q_m}\right)^2 - \frac{1}{2} \left(\sum_\lambda \frac{\partial \alpha_{\lambda\mu}}{\partial Q_m}\right)^2}. \quad (3)$$

Indices $\lambda$ and $\mu$ design the axis of the molecular reference frame $X, Y, Z$. So, to calculated the intensity of some mode of DNA conformational vibrations by the formula (2) the frequencies of vibrations, amplitudes of vibrations, and the derivatives of polarizability tensor with respect to the coordinates of normal vibrations should be determined.

Starting with the known ratio between generalized and normal coordinates of vibrations [24], the derivatives of polarizability tensor with respect to the coordinates of normal vibrations may be written in the following form:

$$\frac{\partial \alpha_{\lambda\mu}}{\partial Q_m} Q_{0m} = \sum_n \frac{\partial \alpha_{\lambda\mu}}{\partial q_n} \cdot \frac{\partial q_n}{\partial Q_m} Q_{0m} = \sum_n \frac{\partial \alpha_{\lambda\mu}}{\partial q_n} \tilde{q}_n^m, \quad (4)$$

where $n$ denotes some mode of the four-mass model. As follows from the formula (4), the product $\partial q_n / \partial Q_m \cdot Q_{0m}$ is equaled to the amplitude of some generalized coordinate $\tilde{q}_n^m$ with the account of the vibrational phase.

Considering the polarizability of nucleotide pair as a sum of polarizability tensors of nucleotides $\alpha = \alpha_1 + \alpha_2$ and taking into account the coordinates of the four-mass model, the formula (4) takes the form:

$$\frac{\partial \alpha_{\lambda\mu}}{\partial Q_m} Q_{0m} = \sum_{j=1}^{2} \left[ \left( \frac{\partial \alpha_{j\lambda\mu}}{\partial X_j} + \frac{\partial \alpha_{j+1\lambda\mu}}{\partial X_j} \right) \tilde{X}_j + \left( \frac{\partial \alpha_{j\lambda\mu}}{\partial Y_j} + \frac{\partial \alpha_{j+1\lambda\mu}}{\partial Y_j} \right) \tilde{Y}_j + \left( \frac{\partial \alpha_{j\lambda\mu}}{\partial \theta_j} + \frac{\partial \alpha_{j+1\lambda\mu}}{\partial \theta_j} \right) \tilde{\theta}_j + \left( \frac{\partial \alpha_{j\lambda\mu}}{\partial \rho_j} + \frac{\partial \alpha_{j+1\lambda\mu}}{\partial \rho_j} \right) \tilde{\rho}_j + \left( \frac{\partial \alpha_{j\lambda\mu}}{\partial \xi_j} + \frac{\partial \alpha_{j+1\lambda\mu}}{\partial \xi_j} \right) \tilde{\xi}_j \right], \quad (5)$$

where $j=1,2$ enumerates the chain of the double helix.

In the present work to determine the intensity of DNA conformational vibrations in the Raman spectra zero order approximation of the valence-optic theory is used. In framework of this approximation the influence of neighboring chemical bonds is neglected. Therefore, the derivatives like $\partial \alpha_{j+1,\lambda\mu} / \partial q_j$ in the formula (5) are neglected. The vibrations along the coordinates $X_j$ and $Y_j$ characterize the motions of nucleotides along respective directions as whole. During these motions the lengths and angles of chemical bonds do not change. Therefore, in framework of zero order approximation of valence-optic theory the polarizability of monomer link of the double helix does not change and the derivatives $\partial \alpha_{j\lambda\mu} / \partial X_j$ and $\partial \alpha_{j\lambda\mu} / \partial Y_j$ in (5) may be putted to zero. The contribution of counterion vibrations to the polarizaility changes of nucleotide is evaluated in framework of Silberstain theory [25]. Our estimations show that the contribution of counterions is not essential comparing to the other contributions. Therefore, in zero order approximation the derivatives $\partial \alpha_{j\lambda\mu} / \partial \xi_j$ in (5) are also putted to zero. The phosphate groups do not vibrate along the coordinates $\theta_j$ and $\rho_j$, therefore their contribution of is not considered. As a result the polarizability tensor of nucleotide pair depends on nucleoside mobility mostly, and the formula (5) may be written as follows:

$$\frac{\partial \alpha_{\lambda\mu}}{\partial Q_m} Q_{0m} \approx \sum_{j=1}^{2} \left( \frac{\partial \beta_{j\lambda\mu}}{\partial \theta_j} \tilde{\theta}_j + \frac{\partial \beta_{j\lambda\mu}}{\partial \rho_j} \tilde{\rho}_j \right), \quad (6)$$

where $\beta_{j\lambda\mu}$ are the components of nucleoside polarizability tensor in the molecular reference frame.

The derivatives of nucleoside polarizability tensor with respect to the coordinates of the four-mass model may be determined in two ways. The first way is the numerical determination of nucleoside polarizability tensors for their different conformations. This approach may be performed when the motion trajectories for each chemical bond are known. The second way is the modeling of analytical dependence of polarizability tensor on coordinates allowing determining the derivatives explicitly. In the present work the second approach is used, since it gives the possibility to obtain the analytical form for the intensities of vibrational modes.

To solve this problem let us determine the polarizability tensors of nucleosides in their own reference frames and than reduce them to the molecular reference frame that is general for both nucleosides. In the nucleoside reference frame the axis $x_j$ and $y_j$ are in plane of the nucleotide pair, and the axis $z_j$ is directed along the helix (Fig. 2). The axis $y_j$ is directed from the atom $C_{3'}$ to the mass center of nucleoside. According to the valence-optic theory the polarizability tensor of nucleoside may be presented as a sum of polarizability tensors of chemical bonds determined in common reference frame. Therefore, to find such sum the reference frames of each chemical bond are reduced to the reference frame of nucleoside.

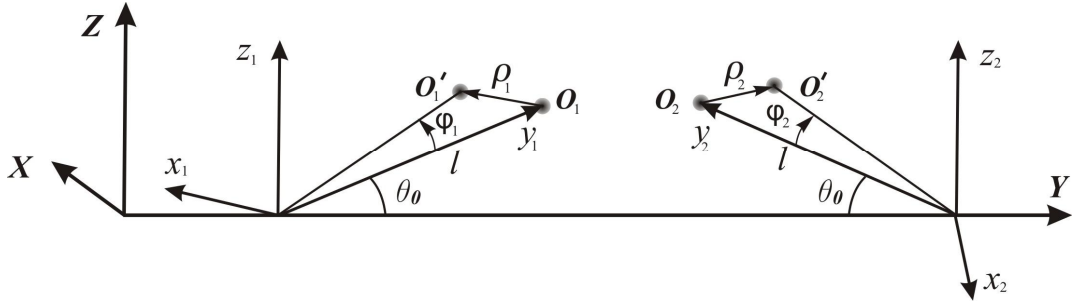

**Fig. 2.** Molecular reference frame (*XYZ*) and reference frames of nucleosides ($x_1y_1z_1$ and $x_2y_2z_2$). The angles $\varphi_1$ and $\varphi_2$ characterize the average angles of displacements of the first and the second nucleosides during the deformations of nucleosides along coordinates $\rho_1$ and $\rho_2$, respectively. The points $O_1$, $O_1'$ and $O_2$, $O_2'$ are the mass centers of the first and the second nucleosides before and after displacement.

The rotation of some bond *i* of nucleoside is performed by the matrix of rotation $\mathbf{P}_i(\theta_j, \rho_j)$ depending on coordinates of the four-mass model. As a result the nucleoside polarizability tensor in its own reference frame has the following form:

$$\boldsymbol{\beta}_j^{nuc}(\theta_j, \rho_j) = \sum_i \mathbf{P}_i(\theta_j, \rho_j) \boldsymbol{\beta}_{ij} \mathbf{P}_i^{-1}(\theta_j, \rho_j), \qquad (7)$$

where *i* enumerates the bonds of nucleoside; $\boldsymbol{\beta}_{ij}$ is the polarizability tensor of *i*-th chemical bond of *j*-th nucleoside in the system of coordinate that is connected with the bond. The reference frame of bond polarizability tensor $\boldsymbol{\beta}_{ij}$ is reduced from its equilibrium position to the nucleoside reference frame using the matrixes of rotation $\mathbf{G}_{ij}$. The deviations of the chemical bonds from equilibrium positions during the conformational vibrations are taken into consideration by means of two matrixes of rotation. One of them is the matrix $\boldsymbol{\Theta}_{ij}(\theta_j)$ describing the deviations from the equilibrium position along the coordinated $\theta_j$, and the second matrix is $\boldsymbol{\Phi}_{ij}(\rho_j)$ describing the displacements along coordinates $\rho_j$. As a result the rotation matrix for some chemical bond has the form of a product of three matrixes:

$$\mathbf{P}_i(\theta_j, \rho_j) = \boldsymbol{\Phi}_{ij}(\rho_j) \boldsymbol{\Theta}_{ij}(\theta_j) \mathbf{G}_{ij}. \qquad (8)$$

Taking into consideration (8) the nucleoside polarizability tensor (7) is obtained:

$$\boldsymbol{\beta}_j^{nuc}(\theta_j, \rho_j) = \sum_i \boldsymbol{\Phi}_{ij}(\rho_j) \boldsymbol{\Theta}_{ij}(\theta_j) \mathbf{G}_{ij} \boldsymbol{\beta}_{ij} \mathbf{G}_{ij}^{-1} \boldsymbol{\Theta}_{ij}^{-1}(\theta_j) \boldsymbol{\Phi}_{ij}^{-1}(\rho_j). \qquad (9)$$

In the equilibrium position the $\boldsymbol{\Phi}_{ij}(\rho_j)$ and $\boldsymbol{\Theta}_{ij}(\theta_j)$ become the unitary matrixes, and the polarizability tensor (9) will be determined by the matrix of rotation angles $\mathbf{G}_{ij}$. These matrixes are individual for each chemical bond in nucleoside and may be determined from coordinates of atoms in nucleoside.

The chemical bonds in DNA structural elements deviate from the equilibrium positions for the same angles, since the conformational vibrations of the double helix are characterized by the motions of whole atomic groups. Therefore, in the case of vibrations of nucleosides as the physical pendulums all chemical bonds deviate from their equilibrium angle $\theta_0$ for the angle $\theta_j$. In the case of intranucleoside vibrations the nucleic base move as a whole for some angle and desoxyribose become deformed. Taking into consideration that the contribution of desoxyribose polarizability to the total polarizability is much lower than the contribution of nucleic bases the deviation angle for all chemical bonds in nucleoside is considered equaled to $\varphi_j$. As a result the formula (9) may be written as follows:

$$\boldsymbol{\beta}_j^{nuc}(\theta_j, \rho_j) \approx \boldsymbol{\Phi}_j(\rho_j) \boldsymbol{\Theta}_j(\theta_j) \boldsymbol{\beta}_{0j}^{nuc} \boldsymbol{\Theta}_j^{-1}(\theta_j) \boldsymbol{\Phi}_j^{-1}(\rho_j), \qquad (10)$$

where matrixes $\boldsymbol{\Phi}_1(\rho_1)$ and $\boldsymbol{\Theta}_1(\theta_1)$ rotate the reference frame clockwise for the angles $\varphi_1$ and $\theta_1$, respectively, and the matrixes $\boldsymbol{\Phi}_2(\rho_2)$ and $\boldsymbol{\Theta}_2(\theta_2)$ rotate the reference frame anticlockwise for the angles $\varphi_2$ and $\theta_2$, respectively (Fig. 2). $\boldsymbol{\beta}_{0j}^{nuc}$ is the equilibrium tensor of nucleoside polarizability in its own reference frame ($x_j y_j z_j$), and in the explicit form it may be written as follows:

$$\boldsymbol{\beta}_{0j}^{nuc} = \sum_i \boldsymbol{G}_{ij} \boldsymbol{\beta}_{ij} \boldsymbol{G}_{ij}^{-1} = \begin{bmatrix} \beta_{jxx} & \beta_{jxy} \\ \beta_{jyx} & \beta_{jyy} \end{bmatrix}. \qquad (11)$$

Note, the conformational vibrations of DNA at the frequency range lower than 200 cm$^{-1}$ are conditioned by the transverse mobility of the structural elements of the double helix [12-14,26]. Taking this into consideration only the components of nucleoside polarizability tensor lying in the plane orthogonal to the helical axis (*XY*-plane) are considered.

To reduce the polarizability tensor to the molecular reference frame (*XYZ*) the reference frame of the first nucleoside ($x_1 y_1 z_1$) should be turned for the angle $\theta_0$ clockwise, and the reference frame of the second nucleoside should be turned for the angle $\pi - \theta_0$ anticlockwise (Fig. 2). Taking into consideration these rotations by matrixes $\boldsymbol{\Theta}_{0j}$ the nucleoside polarizability tensor (10) may be reduced to the molecular reference frame as follows:

$$\boldsymbol{\beta}_j(\theta_j, \rho_j) \approx \boldsymbol{\Theta}_{0j} \boldsymbol{\Phi}_j(\rho_j) \boldsymbol{\Theta}_j(\theta_j) \boldsymbol{\beta}_{0j}^{nuc} \boldsymbol{\Theta}_j^{-1}(\theta_j) \boldsymbol{\Phi}_j^{-1}(\rho_j) \boldsymbol{\Theta}_{0j}^{-1}. \qquad (12)$$

In the case of the first nucleoside the product of matrixes $\boldsymbol{\Theta}_{0j} \boldsymbol{\Phi}_j(\rho_j) \boldsymbol{\Theta}_j(\theta_j)$ turns the reference frame of nucleoside for the angle $\theta_0 + \theta_1 + \varphi_1$ clockwise, and in the case of the second nucleoside for the angle $\pi - \theta_0 - \theta_2 - \varphi_2$ anticlockwise. Taking into consideration the formula (11) the following formula for the nucleoside polarizability in molecular reference frame is obtained:

$$\boldsymbol{\beta}_j = \begin{bmatrix} \cos\Omega_j & (-1)^j \sin\Omega_j \\ (-1)^{j+1} \sin\Omega_j & \cos\Omega_j \end{bmatrix} \begin{bmatrix} \beta_{jxx} & \beta_{jxy} \\ \beta_{jyx} & \beta_{jyy} \end{bmatrix} \begin{bmatrix} \cos\Omega_j & (-1)^{j+1} \sin\Omega_j \\ (-1)^j \sin\Omega_j & \cos\Omega_j \end{bmatrix}, \qquad (13)$$

where $\Omega_j = \theta_0 + \theta_j + \varphi_j$.

According to the developed approach the intranucleoside mobility is considered as the rotation of nucleoside for the angle $\varphi_j$ that is connected with the displacement of the mass center for the distance $\rho_j$ (Fig. 2). Taking into consideration that the displacements from the equilibrium position are small the following correlation may be written:

$$\varphi_j \approx \frac{\rho_j}{l}. \qquad (14)$$

Multiplying the matrixes in (13) and taking into consideration (14), the components of nucleoside polarizability tensors are obtained as functions of coordinates of the four-mass model $\theta_j$ and $\rho_j$:

$$\beta_{jXX} = \beta_{jxx}\cos^2\left(\theta_0+\theta_j+\frac{\rho_j}{l}\right)+(-1)^j\beta_{jxy}\sin 2\left(\theta_0+\theta_j+\frac{\rho_j}{l}\right)+\beta_{jyy}\sin^2\left(\theta_0+\theta_j+\frac{\rho_j}{l}\right);$$

$$\beta_{jXY} = \left(\beta_{jyy}-\beta_{jxx}\right)\frac{(-1)^j}{2}\sin 2\left(\theta_0+\theta_j+\frac{\rho_j}{l}\right)+\beta_{jxy}\cos 2\left(\theta_0+\theta_j+\frac{\rho_j}{l}\right); \qquad (15)$$

$$\beta_{jYY} = \beta_{jxx}\sin^2\left(\theta_0+\theta_j+\frac{\rho_j}{l}\right)-(-1)^j\beta_{jxy}\sin 2\left(\theta_0+\theta_j+\frac{\rho_j}{l}\right)+\beta_{jyy}\cos^2\left(\theta_0+\theta_j+\frac{\rho_j}{l}\right).$$

As a result the derivatives of components of polarizability tensor have the form:

$$\left.\frac{\partial\beta_{jXX}}{\partial\theta_j}\right|_0 = \left(\beta_{jyy}-\beta_{jxx}\right)\sin 2\theta_0+2(-1)^j\beta_{jxy}\cos 2\theta_0;$$

$$\left.\frac{\partial\beta_{jXY}}{\partial\theta_j}\right|_0 = \left(\beta_{jyy}-\beta_{jxx}\right)(-1)^j\cos 2\theta_0-2\beta_{jxy}\sin 2\theta_0; \qquad (16)$$

$$\left.\frac{\partial\beta_{jYY}}{\partial\theta_j}\right|_0 = -\left.\frac{\partial\beta_{jXX}}{\partial\theta_j}\right|_0;$$

$$\left.\frac{\partial\beta_{j\lambda\mu}}{\partial\rho_j}\right|_0 = \frac{1}{l}\left.\frac{\partial\beta_{j\lambda\mu}}{\partial\theta_j}\right|_0.$$

From the formulae (16) follows that the coefficient $A_m$ in (2) is equaled to zero. That is in accordance with known rule for the intensities of deformation vibrations [20], according to which the trace of tensor of derivatives of nucleoside polarizability should be equaled to zero in the case of deformation vibrations.

The product of anisotropy of tensor of derivatives of nucleoside polarizability $B_m$ and the amplitude of normal vibration $Q_{0m}$ may be written as follows:

$$B_m^2 Q_{0m}^2 = 3\left[a_1\left(\tilde{\theta}_1+\frac{\tilde{\rho}_1}{l}\right)_m + s_2\left(\tilde{\theta}_2+\frac{\tilde{\rho}_2}{l}\right)_m\right]^2 + 3\left[s_1\left(\tilde{\theta}_1+\frac{\tilde{\rho}_1}{l}\right)_m - a_2\left(\tilde{\theta}_2+\frac{\tilde{\rho}_2}{l}\right)_m\right]^2, (17)$$

where the coefficients $a_1, s_2, s_1$, and $a_2$ are the combinations of the components of polarizability tensor in the nucleoside reference frame:

$$a_1 = \left(\beta_{1yy}-\beta_{1xx}\right)\sin 2\theta_0 - 2\beta_{1xy}\cos 2\theta_0; \qquad a_2 = \left(\beta_{2yy}-\beta_{2xx}\right)\cos 2\theta_0 - 2\beta_{2xy}\sin 2\theta_0;$$
$$s_2 = \left(\beta_{2yy}-\beta_{2xx}\right)\sin 2\theta_0 + 2\beta_{2xy}\cos 2\theta_0; \qquad s_1 = \left(\beta_{1yy}-\beta_{1xx}\right)\cos 2\theta_0 + 2\beta_{2xy}\sin 2\theta_0; \qquad (18)$$

Taking into consideration the correlation between amplitudes of vibrations for the first and the second nucleosides (1) and the formulae (17), the formula (2) for the intensity of DNA conformational vibrations in the low-frequency Raman spectra may be written in the following form:

$$J_m = J_0 13\frac{2^8\pi^5(\nu_0-\nu_m)^4}{3c^4}\left[\left(a_1\pm s_2\right)^2+\left(s_1\mp a_2\right)^2\right]\cdot\left(\tilde{\theta}+\frac{\tilde{\rho}}{l}\right)_m^2\left(1-e^{-\frac{h\nu_m}{kT}}\right)^{-1}, \qquad (19)$$

where index $m$, as usual, denotes the mode of normal vibrations. In square brackets of the formula (19) upper sing is true for symmetric modes ($\omega_{Ion}^+, \omega_H^+, \omega_{HS}^+$, and $\omega_B^+$) and down sing is true for antisymmetric modes ($\omega_{Ion}^-, \omega_S^-$, and $\omega_B^-$).

## 4. Theory and experiment

The values of frequencies and amplitudes of vibrations for Na- and Cs-DNA are shown in the table 1. Since the amplitudes of vibrations of the first and the second nucleotides are connected by the formula (1), only the amplitudes of vibrations of the first nucleoside are shown. The obtained frequencies of vibrations show that the modes of ion-phosphate vibrations $\omega_{Ion}^+$ and $\omega_{Ion}^-$ for Na-DNA are degenerated the same as in our previous work [17]. In the case of Cs-DNA there is no degeneration of the ion-phosphate modes because in the case of heavy counterions symmetric ion-phosphate vibrations occur with the stretching of H-bonds in the base pairs, and antisymmetric ion-phosphate vibrations do no have H-bond stretching.

The modes of internal conformational vibrations of DNA ($\omega_H^+$, $\omega_S^-$, $\omega_{HS}^+$, $\omega_B^-$, and $\omega_B^+$) are at frequency range from 10 to 110 cm$^{-1}$. Their frequencies in the case of Na-DNA are close to the values that have been obtained earlier in framework of the four-mass model without counterions [12-14]. The increase of the counterion mass induces the softening of the modes of DNA conformational vibrations. Therefore, in the case of Cs-DNA the frequencies of H-bond stretching in the base pairs ($\omega_H^+$ and $\omega_{HS}^+$) and the frequency of intranucleoside vibrations ($\omega_S^-$) decrease for about 15 cm$^{-1}$, and the modes of pendulum-nucleoside vibrations ($\omega_B^-$ and $\omega_B^+$) decrease for about 3 cm$^{-1}$.

**Table 1.** The Frequencies, amplitudes, and Raman intensities for the modes of B-DNA with Na$^+$ and Cs$^+$ counterions. Intensity values are normalized by the intensity of the mode $\omega_{Ion}^+$ of Na-DNA.

| Mode | $\omega_{Ion}^+$ | | $\omega_{Ion}^-$ | | $\omega_H^+$ | | $\omega_S^-$ | | $\omega_{HS}^+$ | | $\omega_B^-$ | | $\omega_B^+$ | |
|---|---|---|---|---|---|---|---|---|---|---|---|---|---|---|
| Ion | Na$^+$ | Cs$^+$ | Na$^+$ | Cs$^+$ | Na$^+$ | Cs$^+$ | Na$^+$ | Cs$^+$ | Na$^+$ | Cs$^+$ | Na$^+$ | Cs$^+$ | Na$^+$ | Cs$^+$ |
| Frequency (см$^{-1}$) | 182 | 118 | 182 | 108 | 111 | 94 | 79 | 58 | 57 | 42 | 16 | 13 | 15 | 12 |
| $\tilde{X}$ (pm) | 0 | 1 | 0 | 1 | 1 | 0 | -3 | -3 | -5 | -6 | -32 | -24 | -32 | -23 |
| $\tilde{Y}$ (pm) | -2 | -4 | -2 | -6 | -2 | -5 | 6 | 1 | 10 | 6 | -17 | -13 | -29 | -29 |
| $\tilde{\theta}$ (°) | 0,11 | 0,31 | 0,10 | 0,28 | 0,20 | 0,03 | 0,02 | 0,24 | 0,60 | 1,04 | 7,00 | 6,99 | 6,97 | 6,92 |
| $\tilde{\rho}$ (pm) | 2 | 7 | 2 | 7 | 7 | 1 | -12 | -9 | -10 | -9 | 0 | 0 | 1 | 1 |
| $\tilde{\xi}$ (pm) | 11 | 6 | 11 | 9 | -1 | 9 | 2 | 7 | 1 | 6 | 0 | -1 | 0 | -1 |
| $\tilde{\delta}$ (pm) | 0 | 8 | 0 | 0 | 10 | -8 | 0 | 0 | 7 | 5 | 0 | 0 | -1 | -1 |
| $J/J_{Ion}^+$ | 1,0 | 15,9 | 0,1 | 23,4 | 11,1 | 0,4 | 4,0 | 2,0 | 5,6 | 0,0 | 476 | 554 | 3686 | 4543 |

The calculated values of the amplitudes of Na-DNA conformational vibrations show that greatest displacements of counterions with respect to the phosphate groups are in the case of ion-phosphate modes (table 1). In the case of internal modes sodium counterions move with the phosphate groups as the masses rigidly bonded to the phosphate groups. The values of amplitudes of internal modes are close to the values that have been obtained in framework of the four-mass model without counterions [13].

In the case of Cs-DNA the character of conformational vibrations is essentially different. The displacements of counterions become large for all modes. The modes of ion-phosphate vibrations are also characterized by large amplitudes of intranucleoside vibrations and H-bond stretching, therefore it resembles the mode $\omega_H^+$. So, the counterions Cs$^+$ play not only the role of counteractive charges, but also disturb the internal conformational vibrations of the DNA double helix.

To calculate the intensities of the modes of conformational vibrations the components of nucleoside polarizability tensor ($\beta_{jxx}$, $\beta_{jxy}$, and $\beta_{jyy}$) are determined as a sum of polarizability tensors of chemical bonds of nucleoside. The coordinates of DNA atoms are taken from X-ray data [27]. The structural parameters of the four-mass model are taken the same as in [12-17]: $l = 4,9$ Å and $\theta_0 = 28°$. The polarizability values for chemical bonds are taken accordingly to the Le Fevre calculation scheme [28]. To simplify further analysis the average values for purine (adenosine and guanosine) and pyrimidine (thymine and cytosine) nucleic bases are considered (table 2).

**Table 2.** Components of nucleoside polarizability tensor (Å$^3$)

| Polarizability component | Purine, $j=1$ | | | Pyrimidine, $j=2$ | | |
|---|---|---|---|---|---|---|
| | Adenosine | Guanosine | Average | Thymine | Cytosine | Average |
| $\beta_{xx}$ | 22,74 | 24,73 | 23,73 | 19,39 | 16,55 | 17,97 |
| $\beta_{xy}$ | -1,98 | -1,31 | -1,65 | -1,41 | -1,52 | -1,47 |
| $\beta_{yy}$ | 22,17 | 21,75 | 21,96 | 19,21 | 16,79 | 18,00 |

Using determined components of nucleoside polarizability tensor and the values for frequencies and amplitudes of vibrations the Raman intensities of the modes of Na- and Cs-DNA conformational vibrations are calculated by the formula (19). The obtained values are normalized by the intensity of the mode $\omega_{Ion}^+$ of Na-DNA (table 1).

To compare our results with the experimental values of frequencies and intensities the spectra of Na- and Cs-DNA are built (Fig. 3). The halfwidth for all spectra lines is equal to 5 cm$^{-1}$ (in the figure 3 these bands are shown by solid line). To take into consideration the heterogeneity of the structural elements and hydrogen bonds in complementary base pairs the spectra with the halfwidth 15 cm$^{-1}$ for the modes $\omega_{HS}^+$ and $\omega_H^+$ are also built (in the figure 3 the resulted spectra in this case is shown by dashed line). Such parameters describe the best the shape of the experimental spectra at frequency range from 40 to 140 cm$^{-1}$.

The obtained results show that the modes of backbone vibrations $\omega_B^+$ and $\omega_B^-$ in the both Na- and Cs-DNA spectra are the most intensive. As follows from the formula (19) such intensity increase may be due to the temperature factor and large amplitudes of vibrations $\tilde{\theta}$ and $\tilde{\rho}$. The temperature factor for the modes $\omega_B^-$ and $\omega_B^+$ is higher about order than the temperature factor of the other modes of conformational vibrations. Moreover, the modes of backbone vibrations are characterized by large amplitudes of vibrations $\tilde{\theta}$ (about 7°) characterizing the vibrations of nucleosides as a pendulums.

In the Raman experimental spectra the intensity of the lowest mode is also higher than the intensity of the other modes of DNA conformational vibrations [29,30], but the experimental values of intensity is not so high as calculated one. That is because in the developed approach the influence of environment is not taken into consideration explicitly. The influence of environment should decrease the amplitudes and frequencies of vibrations. Really, the experimental Raman spectra of DNA [31] show that the intensities of the lowest modes on depend on conditions of the solution and temperature. The authors in [31] consider that this is due to the changes of decrement for the lowest vibrational modes.

At the frequency range higher than 40 cm$^{-1}$ in the spectra of Na-DNA the mode of H-bond stretching in the base pairs $\omega_H^+$ has the greatest intensity. The second mode of H-bond stretching $\omega_{HS}^+$ and the mode of intranucleoside vibrations $\omega_S^-$ have about twice lower intensity (Fig. 3). In the experimental Raman spectra of Na-DNA at frequency range from 50 to 120 cm$^{-1}$ only one broad band is observed [29,30]. According with our calculations it may arise due to the increase of the halfwidth of spectra lines that characterize the stretching of H-bonds in the base pairs $\omega_{HS}^+$ and $\omega_H^+$. Really, considering the increase of the halfwidththe the resulted spectrum has broad shape (dashed line in Fig. 3).

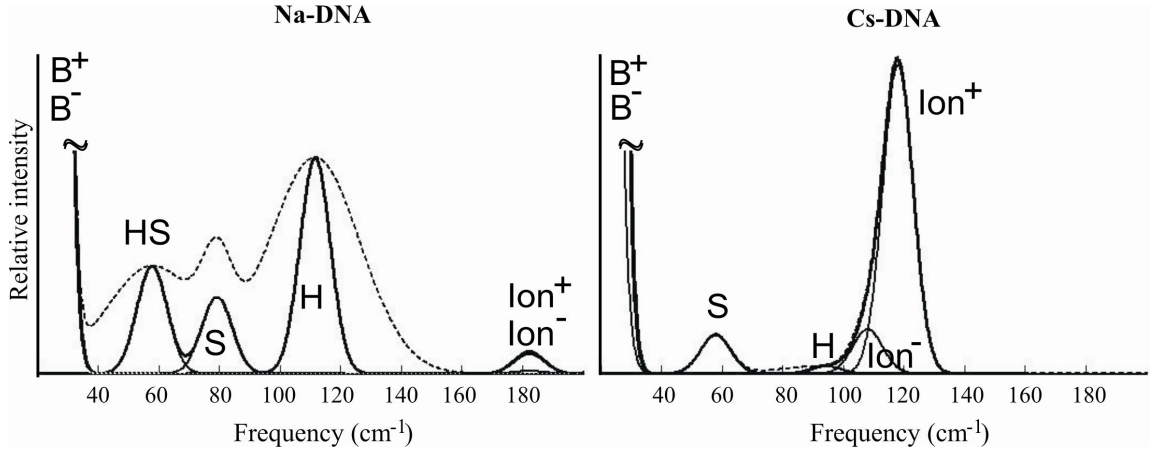

**Fig. 3.** The Raman spectra of Na- and Cs-DNA, built according to the calculation data of the table 1. The halfwidth of the spectra lines is 5 cm$^{-1}$. The dashed line shows the spectra with the halfwidth 15 cm$^{-1}$ for the modes of H-bond stretching in the base pairs $\omega_{HS}^{+}$ and $\omega_{H}^{+}$ (the consideration of heterogeneity).

The calculated spectrum of Na-DNA also agrees with experimental data [19], where the band near 100 cm$^{-1}$ is shown to be consisted of three narrower bands. According to our calculations these bands may be considered as the modes of H-bond stretching in the base pairs $\omega_{HS}^{+}$ (64 cm$^{-1}$), $\omega_{H}^{+}$ (97 cm$^{-1}$) and the mode of intranucleoside vibrations $\omega_{S}^{-}$ (78 cm$^{-1}$). The mode $\omega_{S}^{-}$ has lower intensity than experimental one. However, the calculated spectra shape changes as the halfwidth of the bands $\omega_{HS}^{+}$ and $\omega_{H}^{+}$ increases (dashed line in Fig. 3), and such form of the spectrum is in a good agreement with experimental data [19].

The modes of ion-phosphate vibrations $\omega_{Ion}^{+}$ and $\omega_{Ion}^{-}$ of Na-DNA are at higher frequency range (near 180 cm$^{-1}$), where the broad band describing the vibrations of water molecules is observed experimentally [19]. The calculated values of intensity for the modes of ion-phosphate vibrations are comparatively small (Fig. 3) because of small influence of vibrations of Na$^{+}$ counterions on the vibrations of nucleosides (table 1). This fact explains the reason why the modes of ion-phosphate vibrations do not arise above water band in the spectra of Na-DNA. So, our calculations agree well with the existed experimental data for wet films [29,30] and water solutions of Na-DNA [19].

The calculated spectrum of Cs-DNA shows that the most prominent band is 115 cm$^{-1}$ (Fig. 3). It characterizes the ion-phosphate vibrations $\omega_{Ion}^{+}$ and $\omega_{Ion}^{-}$. The intensity of this band is much higher than the intensity of the respective band in Na-DNA spectra. That is because Cs$^{+}$ counterions influence the amplitudes of intranucleoside vibrations and the vibrations of pendulum-nucleosides (table 1). The intensity of the modes of H-bond stretching in the base pairs $\omega_{H}^{+}$ is lower about order and the intensity of the mode $\omega_{HS}^{+}$ is negligibly small. The intensity of intranucleoside vibrations $\omega_{S}^{-}$ (the band bear 60 cm$^{-1}$) is lower in Cs-DNA than in Na-DNA. Note, in contrast to the spectrum of Na-DNA, the increase of the halfwidth of spectra line for the modes $\omega_{HS}^{+}$ and $\omega_{H}^{+}$ does not change the spectra shape of Cs-DNA. The resulted Raman spectrum of Cs-DNA is more clear than the spectrum of Na-DNA.

The comparison of the calculated spectrum of Cs-DNA with the experimental data [19] show that in the spectrum of Cs-DNA water solutions the narrow band near 100 cm$^{-1}$ is observed. The intensity of this band is about twice higher than the intensity of the bands in Na-DNA. This effect confirms our calculations and indicates that the intensity increase of the band near 100 cm$^{-1}$ in Cs-DNA spectrum is due to the modes of ion-phosphate vibrations.

## 5. Conclusions

Using the developed approach the Raman intensities of the modes of DNA conformational vibrations are calculated. The results for Na- and Cs-DNA show that the modes of backbone vibrations (near 15 cm$^{-1}$) have the greatest intensities that agree with the experimental data. At higher frequency range from 40 to 200 cm$^{-1}$ the modes of H-bond stretching near 60 and 100 cm$^{-1}$ in the base pairs are the most intensive in Na-DNA spectrum. The intensities of other modes are lower. The calculations for Cs-DNA show that at this frequency range the band near 115 cm$^{-1}$ is the most intensive. This band characterizes the ion-phosphate vibrations and its intensity is about twice higher than the intensity of the bands of Na-DNA at frequency range from 60 to 110 cm$^{-1}$. The same difference in Na- and Cs-DNA spectra has been observed experimentally [19]. According to our calculations the modes of ion-phosphate vibrations for Na-DNA are near 180 cm$^{-1}$ and have much lower intensity. So, the modes of ion-phosphate vibrations of Cs-DNA have large intensity in the Raman spectra, while in the case of Na-DNA their intensity are much lower. The developed approach describes the experimental low-frequency Raman spectra of DNA with light and heavy counterions confirming the existence of the modes of ion-phosphate vibrations.

We thank to Prof. M.A. Semenov for challenging discursion of the results of our study.